\documentclass[aps,twocolumn]{revtex4}
\usepackage{bm}
\usepackage{amsmath}
\usepackage{graphicx}
\usepackage{subfigure}
\usepackage[usenames,dvipsnames]{color}
\definecolor{darkblue}{RGB}{0,0,196}
\usepackage[colorlinks=true,linkcolor=darkblue,citecolor=darkblue,urlcolor=darkblue]{hyperref}
\usepackage{setspace}
\usepackage{hyperref}
\usepackage{xcolor}
\hypersetup{
  colorlinks   = true, 
  urlcolor     = red, 
  linkcolor    = blue, 
  citecolor   = blue 
}
\usepackage{footmisc}
\usepackage[makeroom]{cancel}
\usepackage{comment}
\usepackage{lineno}
\def\be{\begin{equation}}
\def\ee{\end{equation}}
\def\ba{\begin{eqnarray}}
\def\ea{\end{eqnarray}}
\usepackage{graphicx}
\usepackage{amsmath,bbm}
\usepackage{amssymb,bm}
\usepackage{yfonts}

\begin{document}
\title{Jet Transport Coefficient at the Large Hadron Collider Energies in a Color String Percolation Approach}

\author{Aditya Nath Mishra$^{1,}$\footnote[1]{e-mail: Aditya.Nath.Mishra@cern.ch}}
\author{Dushmanta Sahu$^{2,}$\footnote[2]{e-mail: Dushmanta.Sahu@cern.ch}}
\author{Raghunath Sahoo$^{2,3,}$\footnote[3]{e-mail: Raghunath.Sahoo@cern.ch (Corresponding Author)}}
\affiliation{$^{1}$Wigner Research Centre for Physics, 29-33 Konkoly-Thege Mikl\'os str., 1121 Budapest, Hungary}
\affiliation{$^{2}$Department of Physics, Indian Institute of Technology Indore, Simrol, Indore 453552, India}
\affiliation{$^{3}$CERN, CH 1211, Geneva 23, Switzerland}

\begin{abstract}
\noindent
Within the color string percolation model (CSPM), jet transport coefficient, $\hat{q}$, is calculated for various multiplicity classes in proton-proton and centrality classes in nucleus-nucleus collisions at the Large Hadron Collider energies for a better understanding of the matter formed in ultra-relativistic collisions. $\hat{q}$ is studied as a function of final state charged particle multiplicity (pseudorapiditydensity
at midrapidity), initial state percolation temperature and energy density. The CSPM results are then compared with different theoretical calculations from the JET Collaboration those incorporate particle energy loss in the medium.

\end{abstract}
\date{\today}
\maketitle 

\section{Introduction}
\label{intro}
The main objective of tera-electron volt energy heavy-ion collisions is to form a Quark-Gluon Plasma (QGP)-- the deconfined state of quarks and gluons, by creating extreme conditions of temperature and(or) energy density~\cite{Bjorken:1982,Enterria:2010}, a scenario that might have been the case after a few microseconds of the creation of the Universe. Jets, collimated emission of a multitude of hadrons originating from the hard partonic scatterings, 
play an important role as hard probes of QGP. These hard jets lose their energy through medium-induced gluon radiation and collisional energy 
loss, as a consequence of which one observes suppression of high transverse momentum particles and the phenomenon is known as jet
 quenching~\cite{Blaizot:1986,Gyulassy:1990,Wang:1992,Baier:1995,Baier:1997,Qin:2015,Gyulassy:2004}. This is a direct signature of a highly
 dense partonic medium, usually formed in high energy heavy-ion collisions. The first evidence of the jet quenching phenomenon has been 
 observed at the Relativistic Heavy-Ion Collider (RHIC)~\cite{Adcox:2001jp,Adler:2002tq,Adcox:2002pe,Adler:2003qi,Adams:2003kv,Adams:2003im,Back:2003qr,Arsene:2003yk,Adare:2006nr,PHENIX:2005,STAR:2005,PHOBOS:2005,BRAHMS:2005,Adare:2008cg} via the measurement of inclusive hadron and jet production at 
 high $p_{\rm{T}}$, $\gamma$-hadron correlation, di-hadron angular correlations and the dijet energy imbalance. The jet quenching phenomena 
 are also widely studied in heavy-ion collisions at the Large Hadron Collider (LHC)~\cite{Aamodt:2010jd,Aad:2010bu,Chatrchyan:2011sx,Chatrchyan:2011pb,Aamodt:2011vg,CMS:2012aa,Chatrchyan:2012nia,Chatrchyan:2012gw,Chatrchyan:2012gt,Aad:2012vca,Chatrchyan:2013exa,Chatrchyan:2013kwa,Chatrchyan:2014ava,Aad:2014wha,Aad:2014bxa}. All the measured observables are found to be strongly modified in central heavy-ion collisions relative to minimum bias proton-proton collisions, when compared to expectations based on treating heavy-ion collisions as an incoherent superposition of independent nucleon-nucleon collisions.

A number of theoretical models that incorporate parton energy loss have been proposed to study the observed jet quenching 
phenomena, namely, Baier-Dokshitzer-Mueller-Peigne-Schiff-Zakharov (BDMPS-Z)~\cite{Zakharov:1996,Baier:1997qr,Baier:1997}, 
Gyulassy-Levai-Vitev (GLV)~\cite{Gyulassy:2000,Gyulassy:2000qr,Gyulassy:2001er} and its CUJET 
implementation~\cite{Buzzatti:2011vt}, high-twist approach (HT-M (Majumder) and HT-BW (Berkeley-Wuhan))~\cite{Guo:2000nz,Wang:2001,Chen:2011vt,Majumder:2009ge,Vitev:2009rd}, Amesto-Salgado-Wiedemann (ASW)~\cite{Wiedemann:2000,Wiedemann:2001}, Arnold-Moore-Yaffe (AMY) model
~\cite{Arnold:2001,Arnold:2002}, MARTINI model~\cite{Schenke:2009gb}, BAMPS model~\cite{Fochler:2010wn}, and LBT model~\cite{He:2015pra}. Most of the theoretical models assumed a static potential for jet-medium interactions, which result in a factorized dependence of parton energy loss on the jet transport coefficient ($\hat{q}$). The jet transport coefficient $\hat{q}$, which describes the average transverse momentum square transferred from the traversing parton, per unit mean free path is a common parameter that modulates the energy loss of jets in a strongly-interacting quantum chromodynamics (QCD) medium~\cite{Baier:1997,Gyulassy:2004}. $\hat{q}$ is also related to the gluon distribution density of the medium and therefore characterizes the medium property as probed by an energetic jet~\cite{Baier:1997,CasalderreySolana:2007sw}. Thus the collision energy and system size dependence study of jet transport coefficient will not only improve our understanding of experimental results on jet quenching but also can directly provide some information about the internal structure of the hot and dense QCD matter~\cite{CasalderreySolana:2007sw,Burke:2013yra}.

In the present work, we study $\hat{q}$ and its relation with various thermodynamic properties of the QCD matter in the framework of the Color String Percolation Model (CSPM)~\cite{nestor,pajares1,Braun2000,pajares2,pajares3,Phyreport} which is inspired by QCD. This can be used as an alternative approach to Color Glass Condensate (CGC) \cite{Phyreport} and is related to the Glasma approach \cite{perx}. In CSPM, it is assumed that color strings are stretched between the projectile and the target, which may decay into new strings via $\it q\bar q$ pair production and subsequently hadronize to produce observed hadrons~\cite{Braun2003}. These color strings may be viewed as small discs in the transverse plane filled with color field created by colliding partons. The final state particles are produced by the Schwinger mechanism, emitting $\it q\bar q$ pairs in this field~\cite{schw}. With the increasing collision energy and size of the colliding nuclei, the number of strings grows and they start interacting to form clusters in the transverse plane. This process is very much similar to discs in the 2-dimensional percolation theory~\cite{epjc71,Isichenko,pajares1,pajares2}. At a certain critical density, called critical percolation density ($\xi_{c}\geq 1.2$), a macroscopic cluster appears that marks the percolation phase transition~\cite{epjc71,Isichenko,pajares1,pajares2,Satz2000,aditya:pos2019}. The combination of the string density dependent cluster formation and the 2-dimensional percolation clustering phase transition are the basic elements of the non-perturbative CSPM. In CSPM, the Schwinger barrier penetration mechanism for particle production and the fluctuations in the associated string tension due to the strong string interactions make it possible to define a temperature. The critical density of percolation is related to the effective critical temperature and thus percolation may provide information on deconfinement in the high-energy collisions~\cite{pajares3,Phyreport}. 
The CSPM approach has been successfully used to describe the initial stages in the soft region in high-energy collisions~\cite{Phyreport,nestor,cunq,andres,epjc71,cpod13,eos2,IS2013,eos3}. 
 In addition to this, CSPM has also been quite successful in estimating various thermodynamic and transport properties of the matter formed in ultra-relativistic energies \cite{Sahoo:2018dcz,Sahoo:2017umy,Sahoo:2019xjq,Sahoo:2018dxn,Sahu:2020nbu,Sahu:2020mzo,Mishra:2020epq}.

The paper runs as follows: first, we present the formulation and methodology of the CSPM approach in section \ref{Formulation}. In section \ref{Results}, we present the result obtained and related discussions. Finally, the important findings of this work are summarized in section \ref{conclusion}.

\section{Formulation and Methodology}
\label{Formulation}
In the color string percolation model, the charged hadron multiplicity, $\mu_{\rm n}$, where $n$ stands for the number of strings in a cluster reduces with the increase of string interactions while the mean transverse momentum squared, $\langle p_{T}^2\rangle_{\rm n}$, of these charged hadrons increases, to conserve the total transverse momentum. The $\mu_{n}$ and $\langle p_{T}^{2}\rangle_{n}$ of the particles produced by a cluster are proportional to the color charge and color field, respectively~\cite{Phyreport,pajares2} and can be defined as

\begin{eqnarray}
	\mu_{n} =\sqrt {\frac {n S_{n}}{S_{1}}}\mu_{1};\hspace{5mm}                                
	\langle p_{t}^{2}\rangle_{n} =\sqrt {\frac {n S_{1}}{S_{n}}} {\langle p_{T}^{2}\rangle_{1}},
	\label{mu_pt1}
\end{eqnarray} 

where $S_{n}$ denotes the transverse overlap area of a cluster of n-strings and the subscript $'1'$ refers to a single string with a transverse overlap area $S_{1} =\pi r_{0}^2$ with the string radius,
$r_{0}$ = 0.2 fm~\cite{Phyreport}, respectively. For the case when strings are just touching each other $S_{n} = n S_{1}$, and $\mu_{n} = n\mu_{1}$, $\langle p_{T}^{2}\rangle_{n}=\langle p_{T}^{2}\rangle_{1}$. When strings fully overlap $S_{n} = S_{1}$  and therefore 
$\mu_{n} =\sqrt{n}\mu_{1}$ and $\langle p_{T}^{2}\rangle_{n}=\sqrt{n}\langle p_{T}^{2}\rangle_{1}$, so that the multiplicity is maximally suppressed and the $\langle p_{T}^{2}\rangle_{n}$ is maximally enhanced. This implies a simple relation between the multiplicity and transverse momentum $\mu_{n}\langle p_{T}^{2}\rangle_{n}=n\mu_{1}\langle p_{T}^{2}\rangle_{1}$, which denotes the conservation of the total transverse momentum. In the thermodynamic limit, one can obtain the average value of $ n S_{1}/S_{n}$ for all the clusters~\cite{pajares1,pajares2} as

\begin{equation}
	\left\langle{n\frac {S_{1}}{S_n}}\right\rangle  =  {\frac {\xi}{1-e^{-\xi}}}\equiv\frac {1}{F(\xi)^{2}}.
	\label{Mean_S1Sn}
\end{equation}

Here, $F(\xi)$ is the color suppression factor by which the overlapping strings reduce the net-color charge of the strings. With $F(\xi)\rightarrow 1$ as $\xi\rightarrow 0$ and $F(\xi)\rightarrow 0$ as $\xi\rightarrow\infty $, where  $\xi =\frac {N_{s} S_{1}}{S_{N}}$ is the percolation density parameter.
Eq.~(\ref{mu_pt1}) can be written as $\mu_{n}=n F(\xi)\mu_{0}$ and 
$\langle p_{T}^{2}\rangle_{n} ={\langle p_{T}^{2}\rangle_{1}}/F(\xi)$.  
It is worth noting that CSPM is a saturation model similar to the Color Glass Condensate (CGC),
where $ {\langle p_{T}^{2}\rangle_{1}}/F(\xi)$ plays the same role as the saturation momentum scale $Q_{s}^{2}$ in the CGC model~\cite{cgc,perx}.

In the present work we have extracted $F(\xi)$ in $pp$ collisions at $\sqrt s$ = 5.02 and 13 TeV for various multiplicity classes using ALICE published results of transverse momentum spectra of charged particles~\cite{alice1}. In case of Pb-Pb collisions at $\sqrt {s_{\rm NN}}$ = 2.76 and 5.02 TeV,~\cite{alice2} and Xe-Xe collisions at $\sqrt {s_{\rm NN}}$ = 5.44 TeV~\cite{alice3}, $F(\xi)$ values have been obtained from the published centrality-dependent transverse momentum spectra of charged particles  by ALICE. To evaluate the initial value of $F(\xi)$ from data, a parameterization of the experimental data of $p_{T}$ distribution in low-energy $pp$ collisions at $\sqrt s$ = 200 GeV (minimum bias), where strings have very low overlap probability, was used~\cite{epjc71}. The $p_{T}$ spectrum of charged particles can be described by a power-law~\cite{Phyreport}:
\begin{eqnarray}
	\frac{d^{2}N_{\rm ch}}{dp_{T}^{2}} =\frac{a}{(p_{0}+{p_{T}})^{\alpha}},
	\label{fitpower}
\end{eqnarray}
where $a$ is the normalisation factor and  $p_{0}$, $\alpha$ are fitting parameters given as, $p_{0}$ = 1.98 and $\alpha$ = 12.87~\cite{Phyreport}. This parameterization is used in high-multiplicity ${\it pp}$ and centrality-dependent heavy-ion (AA) collisions to take into account the interactions of the strings~\cite{Phyreport}. The parameter $p_{0}$ in  Eq.~(\ref{fitpower}) is for independent strings and gets modified to

\begin{equation}
	p_{0}\rightarrow p_{0}\left(\frac {\langle nS_{1}/S_{n}\rangle^{mod}}{\langle nS_{1}/S_{n}\rangle_{pp}}\right)^{1/4}. 
	\label{p0}
\end{equation}

Using Eq.(\ref{p0}) and Eq.(\ref{Mean_S1Sn}) in Eq.(\ref{fitpower}), we get

\begin{eqnarray}
	\frac{d^{2}N_{\rm ch}}{dp_{T}^{2}} =\frac{a}{(p_{0}\sqrt {{F(\xi)_{pp}/F(\xi)}^{mod}}+{p_{T}})^{\alpha}},
	\label{fitpower2}
\end{eqnarray}

where $F(\xi)^{mod}$ is the modified color suppression factor and is used in extracting $F(\xi)$  both in $pp$ and AA collisions. The spectra were fitted using  Eq.(\ref{fitpower2}) in the softer sector with $p_{T}$ in the range 0.15 - 1.0 GeV/c. In $pp$ collisions at low-energies only two strings are exchanged with low probability of interactions, so that
$\langle nS_{1}/S_{n}\rangle_{pp}$ $\approx$ 1, which transforms Eq.(\ref{fitpower2}) into
\begin{equation} 
	\frac{d^{2}N_{ch}}{dp_{T}^{2}} =\frac{a}{(p_{0}\sqrt {1/F(\xi)^{mod}}+{p_{T}})^{\alpha}}.
	\label{spectra2} 
\end{equation}

In the thermodynamic limit, the color suppression factor $F(\xi)$ is related to the percolation density parameter $\xi $ as

\begin{eqnarray}
	F(\xi) =\sqrt\frac{1-e^{-\xi}}{\xi}.
	\label{eq6}
\end{eqnarray}

\section{Results and Discussions}
\label{Results}
In the present work, we have extracted $F(\xi)$ in the multiplicity-dependent $pp$ collisions at $\sqrt{s}$ = 5.02 and 13 TeV~\cite{alice1} and centrality-dependent Pb-Pb collisions at $\sqrt {s_{\rm NN}}$ = 2.76 and 5.02 TeV~\cite{alice2}, and Xe-Xe collisions at $\sqrt {s_{\rm NN}}$ = 5.44 TeV~\cite{alice3} from the charged particles $p_{T}$ spectra measured in ALICE at the LHC. 

\begin{figure}[thbp]
	\centering     
	\vspace*{-0.2cm}
	\includegraphics[scale = 0.44]{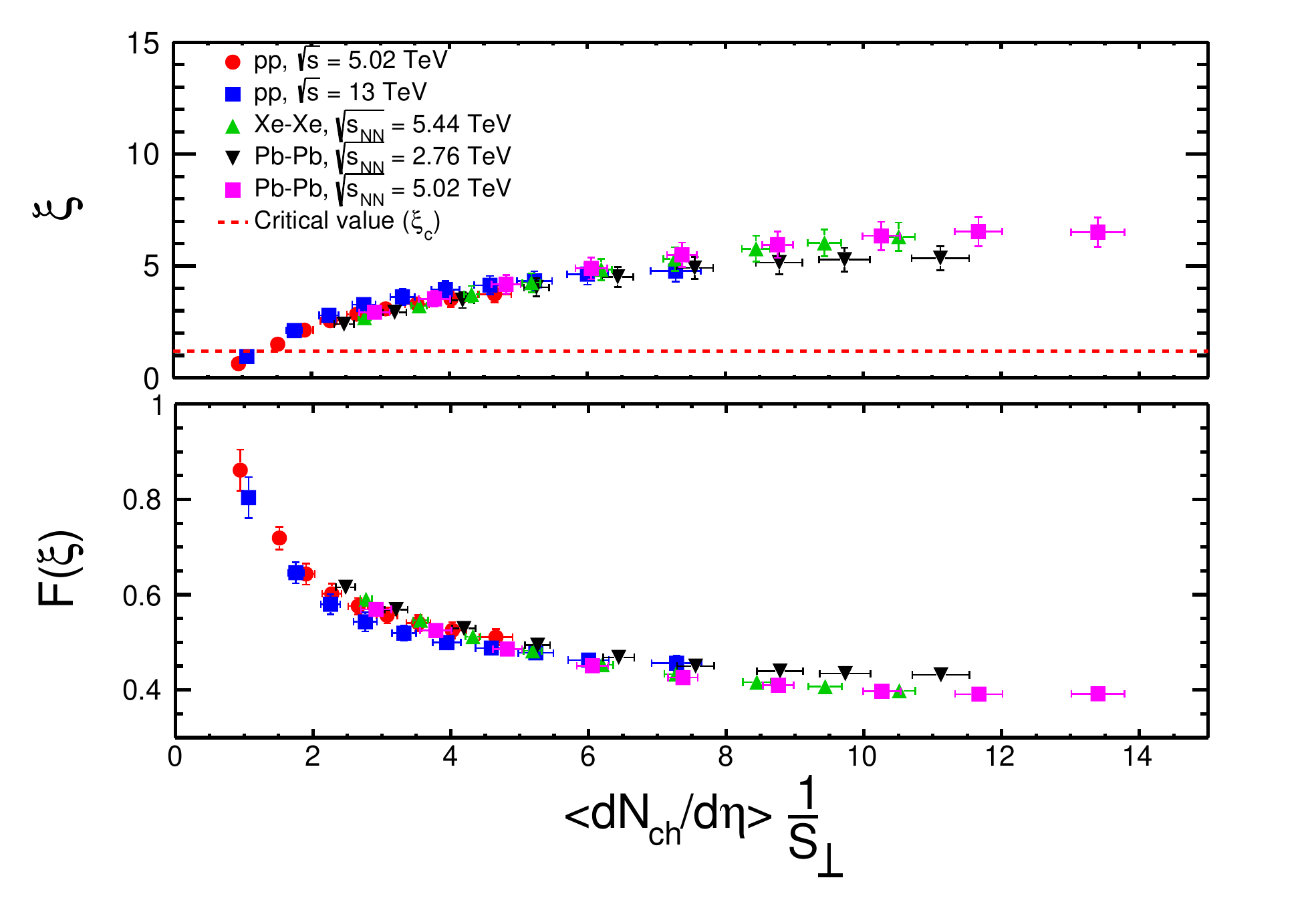}
	\vspace*{-0.5cm}
	\caption{(Color Online) Percolation density parameter, $\xi$, and color suppression factor, $F(\xi)$, as functions of charged particle multiplicity (within $|\eta| < $ 0.8) scaled with the transverse overlap area $S_{\perp}$ in $pp$, Xe-Xe and Pb-Pb collisions are shown in upper and lower panel, respectively. For $pp$ collisions multiplicity-dependent $S_{\perp}$ is obtained from IP-Glasma model~\cite{cross}. In case of Xe-Xe and Pb-Pb collisions we use $S_{\perp}$ values obtained using the Glauber model~\cite{glauber}.}
	\label{fig:fxipp}
\end{figure}  

We show $\xi$ and $F(\xi)$ as functions of final charged particle multiplicity for $pp$, Xe-Xe and Pb-Pb collisions in Fig. \ref{fig:fxipp} upper and lower panel, respectively. The error in $F(\xi)$ is obtained by changing the fitting ranges of the transverse momentum spectra and is found within $\sim$ 3$\%$. For a better comparison of proton-proton and nucleus-nucleus collisions, $ \langle d\rm N_{\rm ch}/d\eta \rangle$ is scaled by the transverse overlap area $S_{\perp}$ for both $pp$ and heavy-ion collisions. For $pp$ collisions, multiplicity-dependent $S_{\perp}$ is calculated using the IP-Glasma model~\cite{cross}. In the case of heavy-ion collisions, the transverse overlap area was obtained using the Glauber model calculations~\cite{glauber}. It is observed that $F(\xi)$ falls onto a universal scaling curve for proton-proton and nucleus-nucleus collisions. Particularly, in the most central heavy-ion collisions (high N$_{tracks}$) and high-multiplicity $pp$ collisions, $F(\xi)$ values fall in a line. This suggests that the color suppression factor is independent of collision energies and collision systems in the domain of high final state multiplicity. Further, what decides the color suppression factor is the final state
multiplicity density of the system, which turns out to be the initial parton density in a system for the case of an isentropic expansion.
\subsection{Temperature}
The connection between $F(\xi)$ and the initial percolation temperature $T(\xi)$ involves the Schwinger mechanism for particle production~\cite{schw, pajares3,Phyreport} and can be expressed as~\cite{pajares3,epjc71}

\begin{equation}
	T(\xi) = {\sqrt {\frac {\langle p_{T}^{2}\rangle_{1}}{ 2 F(\xi)}}}.
	\label{temp}
\end{equation} 
We adopt the point of view that the universal hadronization temperature is a good measure of the upper end of the cross-over phase transition temperature $T_{h}$~\cite{bec1}. The single string average transverse momentum ${\langle p_{T}^{2}\rangle_{1}}$ is calculated at the critical percolation density parameter $\xi_{c}$ = 1.2 with the universal hadronization temperature 
$T_{h}$ = 167.7 $\pm$ 2.6 MeV~\cite{bec1}. This gives\mbox{$\sqrt {\langle {p_{T}^{2}}\rangle _{1}}$ = 207.2 $\pm$ 3.3 MeV}.

\begin{figure}[thbp]
	\centering     
	\vspace*{-0.2cm}
	\includegraphics[scale = 0.44]{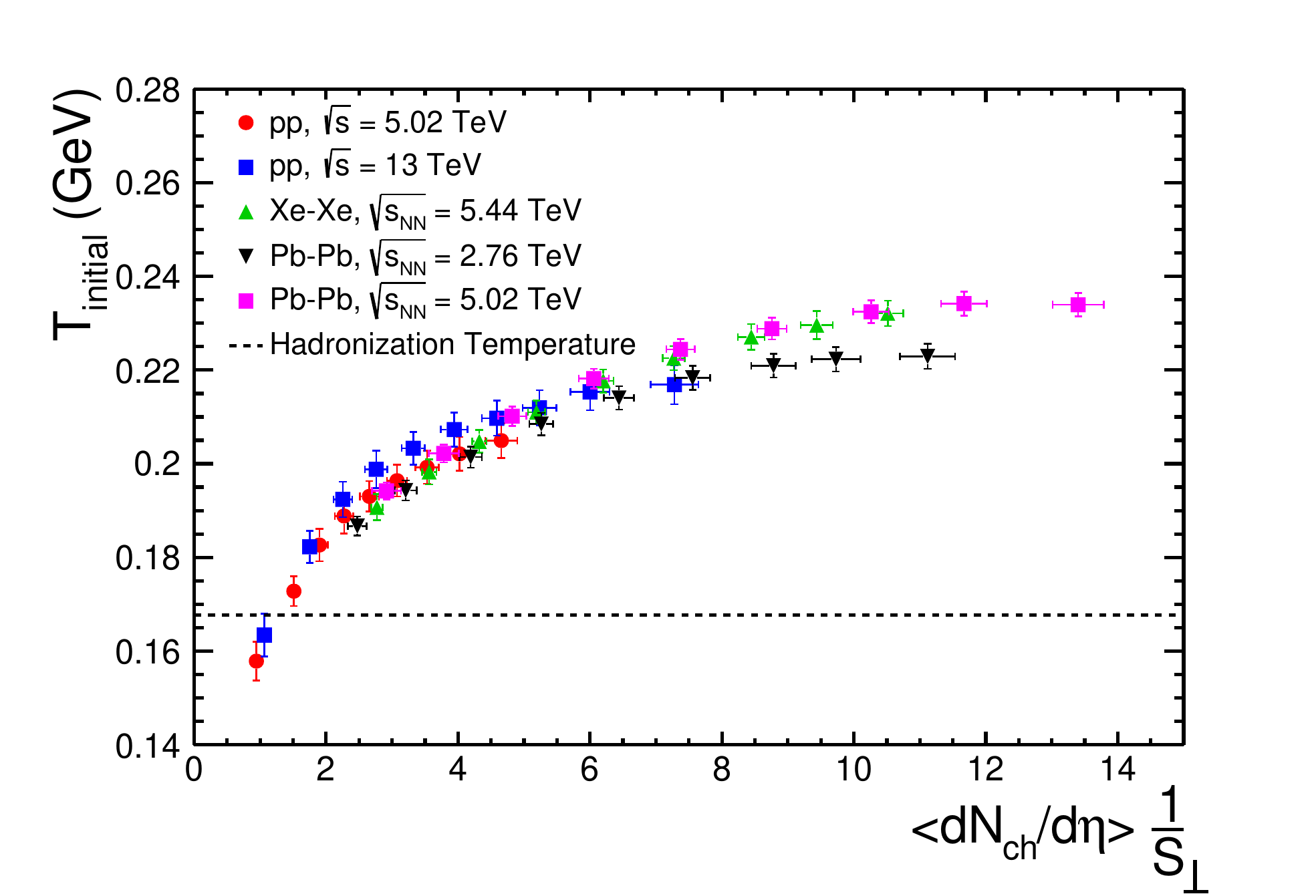}
	\vspace*{-0.5cm}
	\caption{(Color Online) Initial percolation temperature vs $ \langle d\rm N_{\rm ch}/d\eta \rangle$ scaled by $S_{\perp}$ from $pp$, Pb-Pb and Xe-Xe collisions. The line $\sim 167.7$ MeV is the universal hadronization temperature~\cite{bec1}.}
	\label{fig:fxi_temp}
\end{figure}  

In this way at \mbox{$\xi_{c}$ = 1.2} the connectivity percolation transition at $T(\xi_{c})$ models the thermal deconfinement transition. The temperature obtained for most central Pb-Pb collisions at \mbox{$\sqrt{s_{\rm NN}}$ = 2.76 TeV} in this work is $\sim$ 223 MeV, whereas the direct photon measurement up to $p_T < $ 10 GeV/c gives the initial temperature $T_{i}$ = 297 $\pm$ $12^{stat}\pm 41^{sys}$ MeV for 0-20\% central Pb-Pb collisions at \mbox{$\sqrt{s_{\rm NN}}$ = 2.76 TeV} measured by the ALICE Collaboration~\cite{ALICE:2015xmh}. The measured temperature shows that the temperature obtained using Eq.~(\ref{temp}) can be termed as the temperature of the percolation cluster.

Figure~\ref{fig:fxi_temp} shows a plot of initial temperature from CSPM as a function of $ \langle d\rm N_{\rm ch}/d\eta \rangle$ scaled by $S_{\perp}$. Temperatures from both $pp$ and AA collisions fall on a universal curve when multiplicity is scaled by the transverse overlap area. The horizontal line at $\sim $ 167.7 MeV is the universal hadronization temperature obtained from the systematic comparison of the statistical thermal model parametrization of hadron abundances measured in high energy $e^{+}e^{-}$, $pp$ and AA collisions~\cite{bec1}. One can see that temperature for higher multiplicity classes in $pp$ collisions at $\sqrt s$ = 5.02 and 13 TeV, are higher than the hadronization temperature and similar to those observed in Xe-Xe at $\sqrt {s_{\rm NN}}$ = 5.44 TeV and Pb-Pb collisions at $\sqrt {s_{\rm NN}}$ = 2.76 and 5.02 TeV.
\subsection{Energy density}
The calculation of the bulk properties of hot QCD matter and characterization of the nature of the QCD phase transition is one of the most important and fundamental problems in finite-temperature QCD. The QGP, according to CSPM, is born in local thermal equilibrium because the temperature is determined at the string level. Beyond the initial temperature, $T > T_{c}$ the CSPM perfect fluid may expand according to Bjorken boost invariant 1-dimension hydrodynamics~\cite{Bjorken}. In this framework, the initial energy density is given by:

\begin{eqnarray}
	\varepsilon =\frac{3}{2}\frac{\frac{dN_{\rm ch}}{dy}\langle m_{T}\rangle}{S_{\rm N}\tau_{\rm pro}},
	\label{p1}
\end{eqnarray}

where $\varepsilon$ is the energy density, $S_{\rm N}$ is the transverse overlap area and $\tau_{\rm pro}$, the production time for a boson (gluon), is described by~\cite{Wong}

\begin{eqnarray}
	\tau_{\rm pro} =\frac{2.405\hbar}{\langle m_{T}\rangle}.
	\label{tau}
\end{eqnarray}

Here, $m_T$ = $\sqrt{m^2+p_T^2}$ is the transverse mass. For evaluating $\varepsilon$, we use the charged particle multiplicity $dN_{\rm ch}/dy$ at mid-rapidity and $m$ is taken as the pion mass (pions being the most abundant particles in a multiparticle production process like that is discussed here), which gives the lower bound of the energy density. For the estimation of $\langle m_T \rangle$, we use the $p_T$ spectra of pions at different
collision energies and collision species in the $p_T$ range: $0.15 ~{\rm GeV/c} < p_T < 1$ GeV/c.

The purpose of estimating the initial percolation temperature and the initial energy density in the framework of CSPM is to study the jet transport
coefficient as a function of these global observables for different collision species and collision energies at the LHC. Let us now proceed to
estimate $\hat{q}$ in the CSPM framework.

\subsection{Jet transport coefficient}
\label{qhat}
The final state hadrons, produced in ultra-relativistic collisions at large transverse momenta, are strongly suppressed in central collisions compared to peripheral collisions. This suppression of hadrons at high $p_{\rm T}$, which is usually referred to as jet quenching, is believed to be the result of the parton energy loss induced by multiple collisions in the strongly interacting medium. Thus, we are encouraged to study the jet transport coefficient, $\hat{q}$, which encodes the parton energy loss in the medium. It is also related to the $p_{\rm T}$ broadening of the energetic partons propagating inside the medium. In kinetic theory framework, $\hat{q}$ can be estimated by the formula \cite{Baier:2008js},

 \begin{eqnarray}
\hat{q} = \rho \int_{}^{}d^{2}q_{\perp} ~q_{\perp}^{2} \frac{d\sigma}{d^{2}q_{\perp}},
	\label{qhat}
\end{eqnarray}
where $\rho$ is the number density of the constituents of the medium, $q_{\perp}$ is the transverse momentum exchange between the jet and the medium, and $\frac{d\sigma}{d^{2}q_{\perp}}$ denotes the differential scattering cross-section of the particles inside the medium.

The transport parameter of jet quenching, $\hat{q}$, and the shear viscosity-to-entropy density ratio ($\eta/s$), transport parameters describing the exchange of energy and momentum between fast partons and medium, are directly related to each other as~\cite{Liu:2006,Majumder:2007,CasalderreySolana:2007sw,Xu:2013}
\begin{eqnarray}
	\frac{\eta}{s} \approx\frac{3}{2}\frac{T^{3}}{\hat{q}}
	\label{etaByS_qhat}
\end{eqnarray}
Within the CSPM approach, the shear viscosity-to-entropy density ratio, $\eta/s$, can be expressed as \cite{Phyreport,Sahu:2020mzo}
\begin{eqnarray}
	\frac{\eta}{s} =\frac{T L}{5(1 - e^{-\xi})},
	\label{etaByS}
\end{eqnarray}
here $L$ is the longitudinal extension of the string $\sim$ 1 fermi~\cite{pajares3}. 
One can get final expression for jet transport coefficient from Eq. \ref{etaByS_qhat} as:

\begin{eqnarray}
	\hat{q}\approx\frac{3}{2}\frac{T^{3}}{\eta/s}\approx\frac{15}{2}\frac{T^{2} (1 - e^{-\xi})}{L} 
	\label{eq:qhat}
\end{eqnarray}

\begin{figure}[ht!]
\centering
		\includegraphics[scale = 0.44]{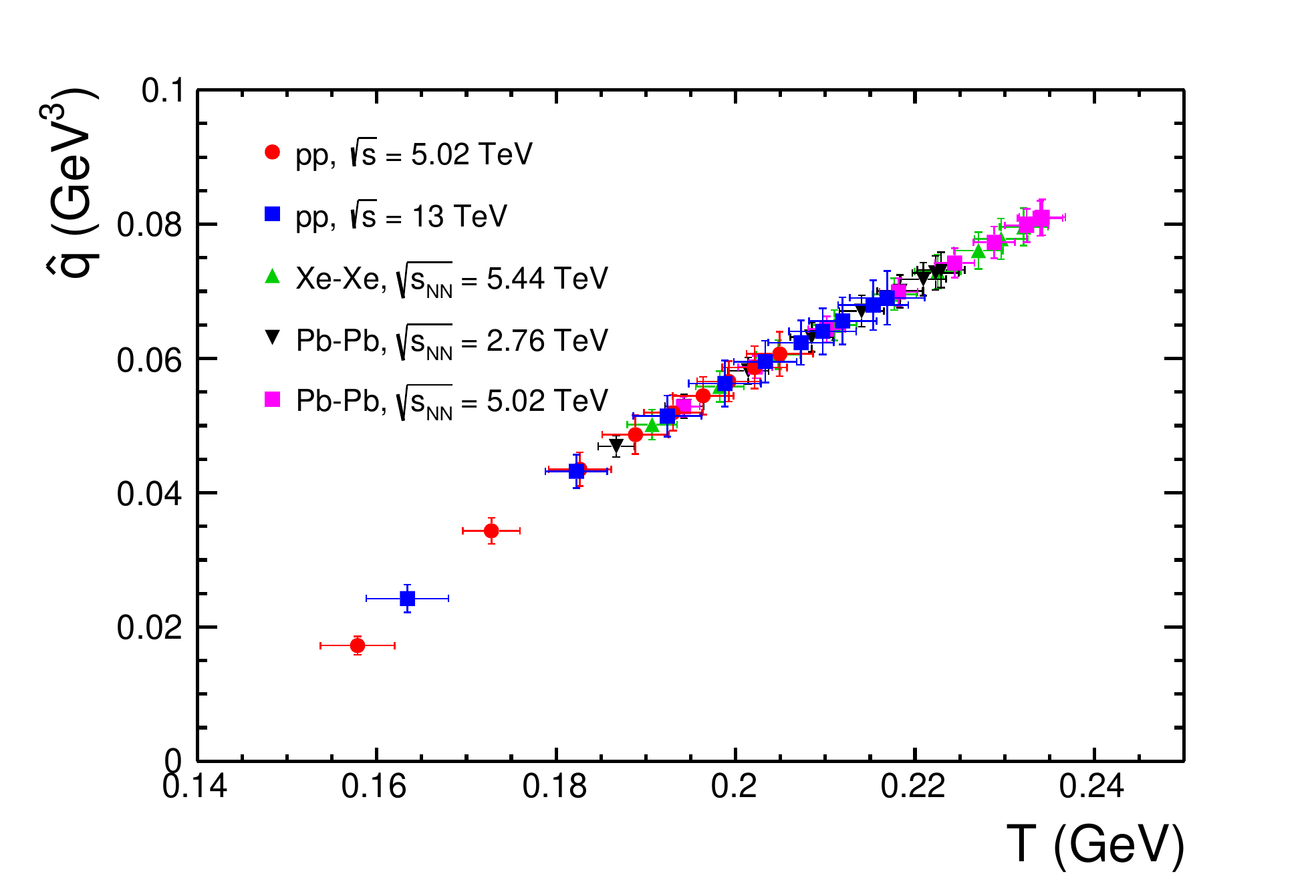}
		\caption{(Color Online) Jet quenching parameter $\hat q$ as a function of temperature within the CSPM for $pp$ collisions at $\sqrt{s}$ = 5.02 and 13 TeV, Xe-Xe collisions at $\sqrt{s_{\rm NN}}$ = 5.44 TeV and Pb-Pb collisions at $\sqrt{s_{\rm NN}}$ = 2.76 and 5.02 TeV.}
		\label{fig:qhatVsT}
\end{figure}

\begin{figure}[ht!]
	\centering
	\includegraphics[scale = 0.44]{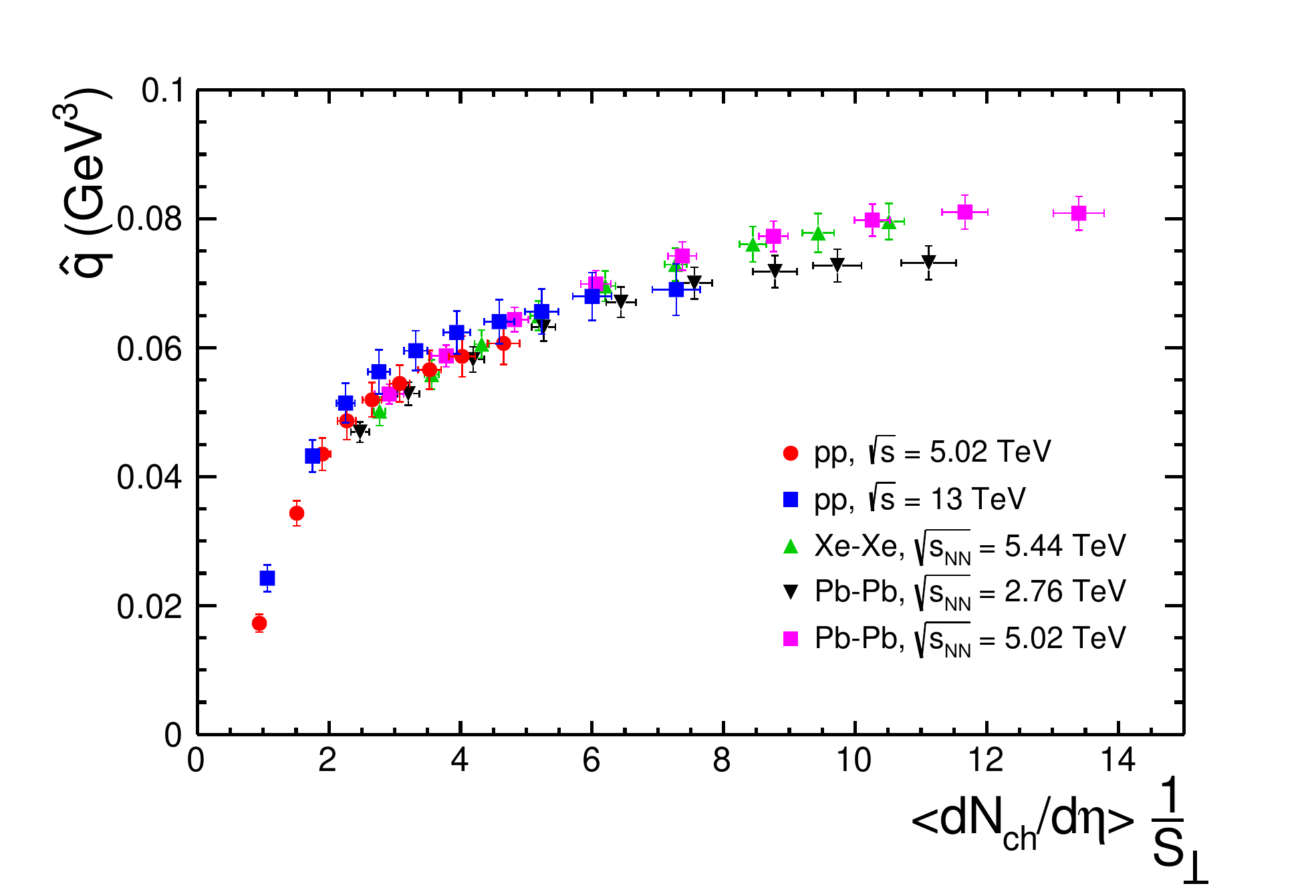}
	\caption{(Color Online) Jet quenching parameter $\hat q$ as a function charged particle multiplicity scaled with transverse overlap area ($S_{\perp}$) within the CSPM for $pp$ collisions at $\sqrt{s}$ = 5.02 and 13 TeV, Xe-Xe collisions at $\sqrt{s_{\rm NN}}$ = 5.44 TeV and Pb-Pb collisions at $\sqrt{s_{\rm NN}}$ = 2.76 and 5.02 TeV.}
	\label{fig:qhatVsNch}
\end{figure}

\begin{figure}[thbp]
	\centering     
	\vspace*{-0.2cm}
	\centerline{\includegraphics[width=9.5cm]{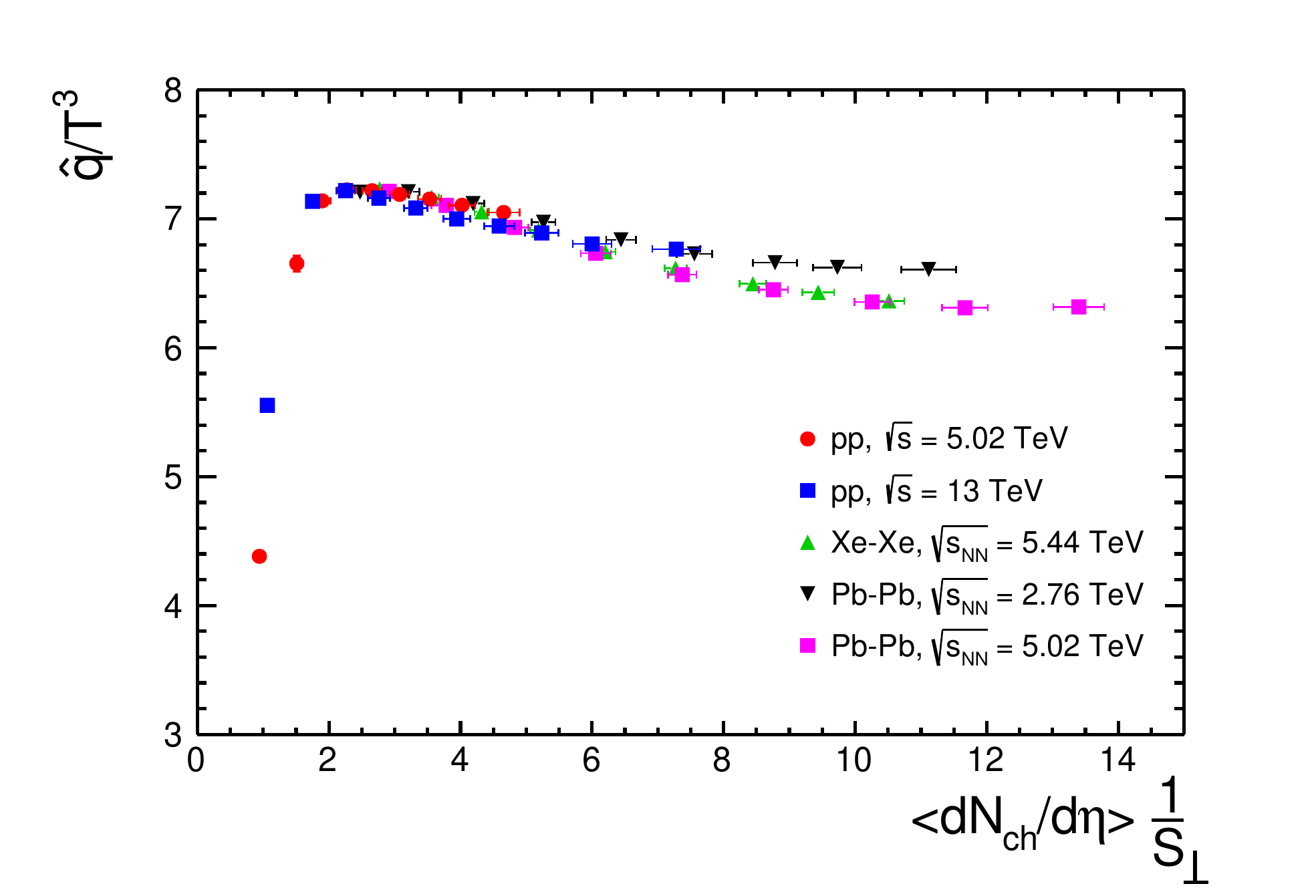}}		
	\vspace*{-0.5cm}
	\caption{(Color Online) $\hat{q}/T^{3}$ vs charged particle multiplicity scaled by $S_{\perp}$ for $pp$ collisions at $\sqrt{s}$ = 5.02 and 13 TeV, Xe-Xe collisions at $\sqrt{s_{\rm NN}}$ = 5.44 TeV and Pb-Pb collisions at $\sqrt{s_{\rm NN}}$ = 2.76 and 5.02 TeV.}
	\label{fig:qhatbyT3VsNch}
\end{figure}  

The jet quenching parameter $\hat q$ is plotted as a function of initial percolation temperature in Fig. \ref{fig:qhatVsT}. Interestingly, we observe a linear increase in $\hat q$, with the increase in temperature for both $pp$ and AA collisions. At low temperatures, the value of jet quenching parameter is around 0.02 $\rm GeV^{3}$. This value increases gradually and at high temperatures, it reaches the value around 0.08 $\rm GeV^{3}$. 

The Jet Collaboration has also extracted $\hat q$ values from five different hydrodynamic approaches with the initial temperatures of 346-373 and 447-486 MeV for the most central Au-Au collisions at $\sqrt{s}=200$ GeV at RHIC and Pb-Pb collisions at $\sqrt{s}=2.76$ TeV at LHC, respectively. The variation of $\hat q$ values between different hydrodynamic models is considered as theoretical uncertainties. The scaled jet quenching parameter $\hat {q}/T^{3}$ at the highest temperatures reached in the most central Au--Au and Pb--Pb collisions are~\cite{Burke:2013yra}
 
\begin{equation*}
	\frac{\hat q}{T^3}\approx \left\{ 
	\begin{array}{l}
		4.6\pm 1.2 \qquad \text{at RHIC}\\
		3.7 \pm 1.4 \qquad \text{at LHC}.
	\end{array}
	\right.
\end{equation*}

The corresponding absolute values for $\hat q$ for a 10 GeV quark jet are,
\begin{equation*}
	\hat q \approx \left\{ 
	\begin{array}{l}
		0.23\pm 0.05 \\
		0.37 \pm 0.13
	\end{array}
	\;\; {\rm GeV}^3 \;\; \text{at} \;\;
	\begin{array}{l}
		\text{T = 346-373 MeV (RHIC)},\\
		\text{T = 447-486 MeV (LHC)},
	\end{array}
	\right.
\end{equation*}

at an initial time $\tau_0=0.6$ fm/$c$. In this work, we use charged particle spectra to calculate $\hat q$ within the CSPM approach, so we can't reach the initial temperature published by the JET Collaboration. Therefore, our $\hat q$ is significantly smaller than the value published by the JET Collaboration for the most central Pb-Pb collisions at $\sqrt{s}$ = 2.76 TeV at the LHC.

In Fig.~\ref{fig:qhatVsNch}, we have plotted $\hat q$ as a function of charged particle multiplicity scaled with transverse overlap area for $pp$ collisions at $\sqrt{s}$ = 5.02 and 13 TeV, Xe-Xe collisions at $\sqrt{s_{\rm NN}}$ = 5.44 TeV and Pb-Pb collisions at $\sqrt{s_{\rm NN}}$ = 2.76 and 5.02 TeV. One can see that $\hat q$ shows a steep increase at lower charged particle multiplicities in $pp$ collisions and gets saturated at very high-multiplicity for all studied energies. This behaviour suggests that at lower multiplicities, the system is not dense enough to highly quench the partonic jets, whereas with the increase of multiplicity, the quenching of jets becomes more prominent.
	
The dimensionless parameter, $T^3$-scaled $\hat q$ is shown in Fig. \ref{fig:qhatbyT3VsNch} as a function of charged particle multiplicity scaled with transverse overlap area. In the low multiplicity regime, we observe a steep increase in $\hat q/T^{3}$, and after reaching a maximum at $ \langle d\rm N_{\rm ch}/d\eta \rangle /S_{\perp} \sim 2 $ it starts decreasing regardless of the collision system or collision energy. The decrease in $\hat q/T^{3}$ is faster in Pb-Pb and Xe-Xe as compared to the $pp$ collisions.

\begin{figure}[ht!]
   \centering
		\includegraphics[scale = 0.44]{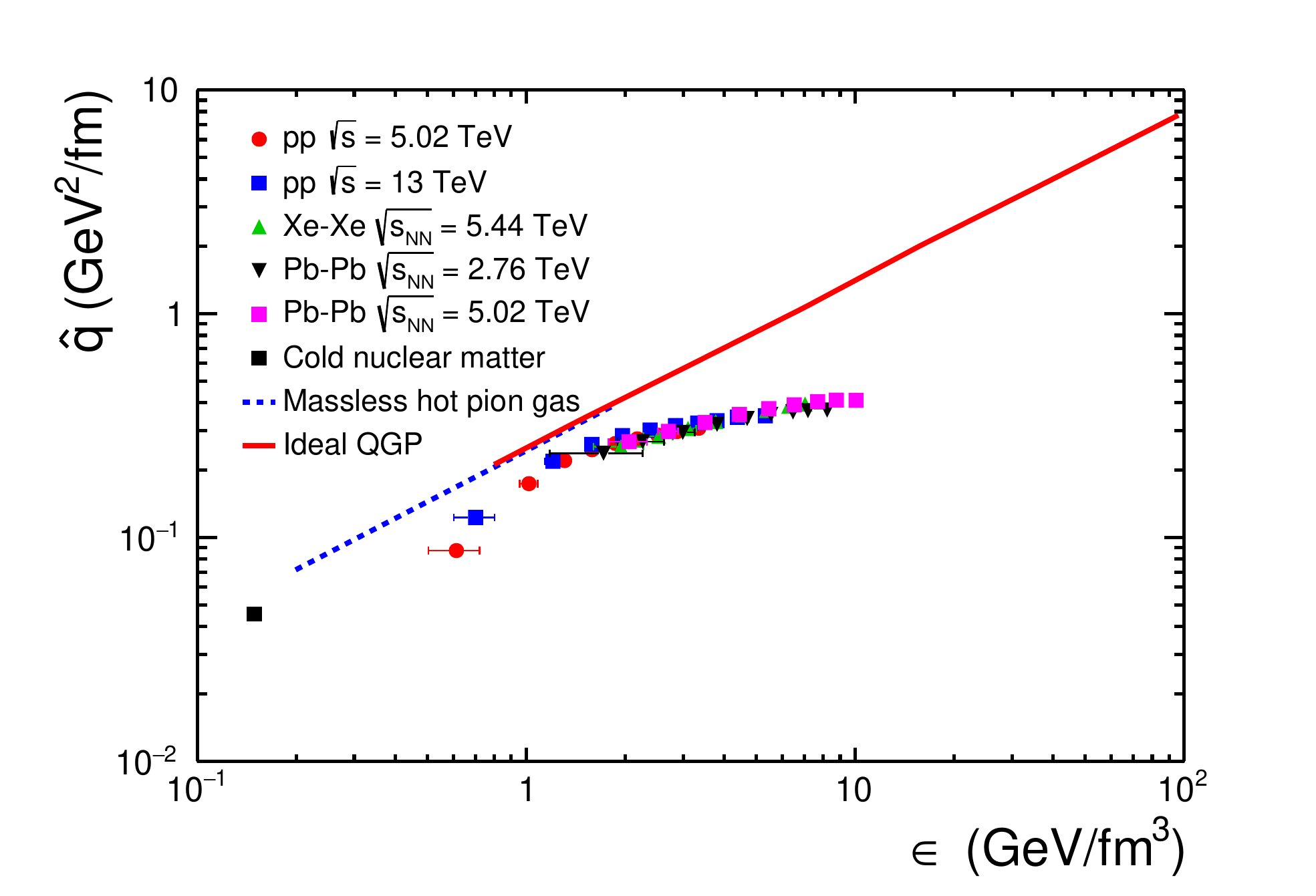}
		\caption{(Color Online) Jet quenching parameter $\hat q$ as a function of initial energy density for $pp$ collisions at $\sqrt{s}$ = 5.02 and 13 TeV, Xe-Xe collisions at $\sqrt{s_{\rm NN}}$ = 5.44 TeV and Pb-Pb collisions at $\sqrt{s_{\rm NN}}$ = 2.76 and 5.02 TeV. The blue dotted line is for massless pion gas, the solid red curve is for ideal QGP and the black square is for cold nuclear matter~\cite{Baier:2003}.}
		\label{fig:qhatVsEdensity}
\end{figure}

 The variation of $\hat q$ as a function of initial energy density is shown in Fig.~\ref{fig:qhatVsEdensity}. To have a better understanding, we have compared our results with that of cold nuclear matter, massless hot pion gas and ideal QGP calculations~\cite{Baier:2003}. We observe that our CSPM result is closer to the massless hot pion gas at low energy density. As initial energy density increases, $\hat q$ values
 increase and then show a saturation towards heavy-ion collisions, which produce a denser medium. The saturation behaviour observed at high energy densities suggests that $\hat q$ remains unaffected after a certain energy density. Similar behaviour is observed when $\hat q$ is studied as a function of multiplicity (shown in Fig.~\ref{fig:qhatVsNch}). The jet energy loss inside a denser QCD medium goes towards
 saturation after a threshold in the final state multiplicity is reached. If we compare the behaviour of $\eta/s$ as a function of $T/T_c$ for
 $T > T_c$ (the domain of validity of CSPM), we observe an increasing trend, which in principle should be reflected in a reverse way
 in the observable $\hat q/T^3$. However, the interplay of higher temperature and lower $\eta/s$ decides the high temperature behavior
 of $\hat q$ as shown in Fig.~\ref{fig:qhatVsEdensity}. Further, one observes the CSPM based estimations of $\hat q$ showing a deviation
 from the ideal QGP behaviour for energy densities higher than 1 GeV/$\rm fm^3$. This is because the ideal QGP calculations of 
 Ref. \cite{Baier:2003}, assumes $\epsilon/T^4$ a constant value, whereas the CSPM-based estimations show an increasing trend of
$\epsilon/T^4$ towards high temperature(energy density or final state multiplicity) \cite{Mishra:2020epq}.

\begin{figure}[ht!]
	\centering
	\includegraphics[scale = 0.44]{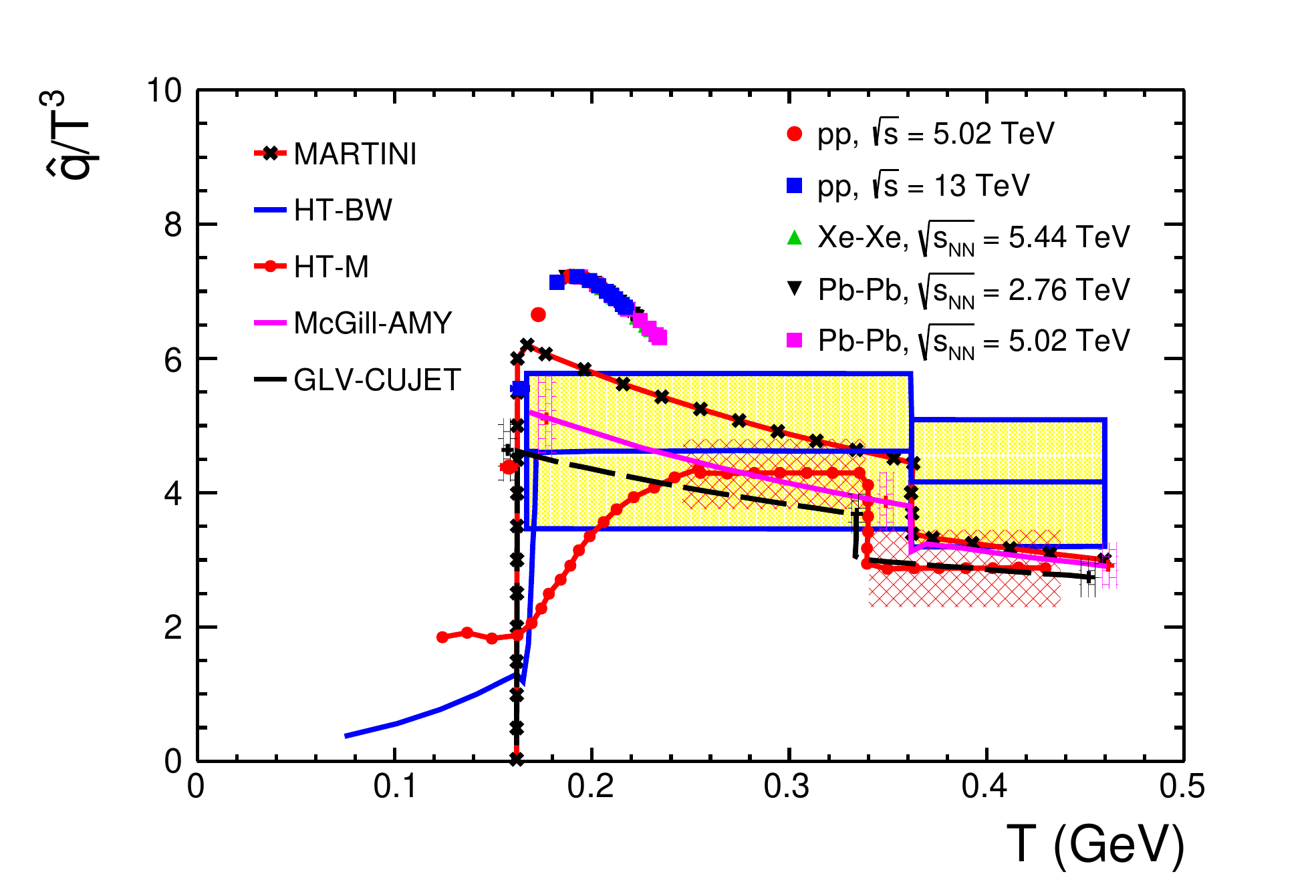}
	\caption{(Color Online) Scaled jet quenching parameter $\hat q/T^{3}$ as a function of initial temperature for $pp$ collisions at $\sqrt{s}$ = 5.02 and 13 TeV, Xe-Xe collisions at $\sqrt{s_{\rm NN}}$ = 5.44 TeV and Pb-Pb collisions at $\sqrt{s_{\rm NN}}$ = 2.76 and 5.02 TeV. The yellow band shows the estimated uncertainties under the high-twist high-twist Berkeley-Wuhan (HT-BW) model, whereas the red shaded region shows the corresponding uncertainty in the high-twist Majumder (HT-M) model. See text for details.}
	\label{fig:qhatByTModel}
\end{figure}

In Fig. \ref{fig:qhatByTModel}, we have plotted $\hat q/T^{3}$ as a function of initial temperature. For the comparison, we have also plotted the results obtained by the JET Collaboration using five different theoretical models that incorporate particle energy loss in the medium. 
The GLV model~\cite{Gyulassy:2000,Gyulassy:2000qr,Gyulassy:2001er} predicted the general form of the evolution of center-of-mass energy of the high transverse momentum pion nuclear modification factor from Super Proton Synchrotron (SPS) and RHIC to LHC energies. CUJET 1.0 explained the similarity between $R_{\rm AA}$ at RHIC and LHC, despite the fact that the initial QGP density in LHC almost doubles that of RHIC, by taking the effects due to multi-scale running of the QCD coupling $\alpha(Q^2)$ into account~\cite{Buzzatti:2011vt}. In CUJET 2.0, the CUJET 1.0 is coupled with the 2+1D viscous hydro fields. By taking GLV-CUJET, the JET Collaboration has estimated the scaled $\hat q$, shown by the dashed black line. The Higher-Twist Berkeley-Wuhan (HT-BW) model uses a 3+1D ideal hydrodynamics to provide the space-time evolution of the local temperature and the flow velocity in the medium along the jet propagation path in heavy-ion collisions. The result obtained from HT-BW model is represented by the blue line. The Higher-Twist Majumder (HT-M) model (red line with filled circles) uses a 2+1D viscous hydrodynamic model to provide the space-time evolution of the entropy density~\cite{Guo:2000nz,Wang:2001,Chen:2011vt,Majumder:2009ge,Vitev:2009rd}. The nuclear initial parton scatterings for jet production are carried out by using PYTHIA8 in the MARTINI model~\cite{Schenke:2009gb}. This model describes the suppression of hadron spectra in heavy-ion collisions at RHIC very well with a fixed value of the strong coupling constant. The MARTINI model calculation is represented by the red line with black crosses. In MCGILL-AMY model~\cite{Arnold:2001,Arnold:2002}, the scattering and radiation processes are described by thermal QCD and hard thermal loop (HTL) effects~\cite{Su:2012} and Landau-Pomeranchuck-Migdal (LPM) interference~\cite{Baier:1996}. In this approach, a set of rate equations for their momentum distributions are solved to obtain the evolution of hard jets (quarks and gluons) in the hot QCD medium. The result obtained from this model is represented by the magenta line. We observe that $\hat q/T^{3}$ obtained from the CSPM approach has a similar kind of behaviour as observed by JET Collaboration.

\section{Conclusion}
\label{conclusion}

We study the thermodynamic and transport properties of the matter formed in $pp$ and AA collisions at LHC energies within the framework of the CSPM. We extract percolation density parameter by fitting transverse momentum spectra within Color String Percolation Model and then calculate initial percolation temperature ($T$), energy density ($\varepsilon$) and the jet transport coefficient ($\hat{q}$). In the present work, for the first time, we study the jet transport coefficient of produced hot QCD matter within the color string percolation approach as a function of
final state charged particle multiplicity at the LHC energies. We show that $\hat q$ increases linearly with initial temperature regardless of the collision system or collision energy. 

At very low multiplicity, $\hat q$ shows a sharp increase and this dependence becomes weak at high-multiplicity (energy density). This behaviour suggests that at lower multiplicity, the system is not dense enough to highly quench the partonic jets, whereas with the increase of multiplicity the quenching of jets becomes more prominent. At very high-multiplicity (energy density), $\hat q$ saturates with multiplicity (energy density). This allows us to conclude that at very high-multiplicity (high energy density), $\hat q$ becomes independent of final state multiplicity when
scaled by the transverse overlap area of the produced fireball. Interestingly, we found that $\hat q$ in the low energy density regime, the system behaves almost like a massless hot pion gas. The $\hat q/T^{3}$ obtained from the CSPM approach as a function of temperature has a similar kind of behaviour as observed by the JET Collaboration using five different theoretical models that incorporate particle energy loss in the medium. 

 In view of heavy-ion-like signatures seen in TeV high-multiplicity $pp$ collisions at the LHC energies, it would be really exciting to see
the jet quenching results in such collisions to infer about the possible QGP-droplet formation. The present study of jet transport coefficient as a
function of final state multiplicity, initial temperature and energy density will pave the way for such an experimental exploration making LHC 
$pp$ collisions unique.

\section*{Acknowledgements}
 ANM thanks the Hungarian National Research, Development and Innovation Office (NKFIH) under the contract numbers OTKA K135515, K123815 and NKFIH 2019-2.1.11-TET-2019-00078, 2019-2.1.11-TET-2019-00050 and Wigner Scientific Computing Laboratory (WSCLAB, the former Wigner GPU Laboratory). R.S. acknowledges the financial supports under the CERN Scientific Associateship and the financial grants under DAE-BRNS Project No. 58/14/29/2019-BRNS of the Government of India.



\begin{thebibliography}{99}	
\bibitem{Bjorken:1982}  J. D. Bjorken. Energy Loss of Energetic Partons in Quark-Gluon Plasma: Possible Extinction of High $p_T$ Jets in Hadron-Hadron Collisions. Report FERMILAB-Pub-82-059-THY. Fermi National Accelerator Laboratory, Batavia, Illinois, USA, 1982. Available online: \url{https://lss.fnal.gov/archive/preprint/fermilab-pub-82-059-t.shtml} (accessed on 20 December 2021).\bibitem{Enterria:2010}  D. d\'Enterria. Jet quenching. In \emph{Landolt-B\"ornstein -- Group. Elementary Particles, Nuclei and Atoms, 
23: Relativistic Heavy Ion Physics}; Stock, R., Ed.; Springer-Verlag, Berlin/Heidelberg, Germany, 2010. 
\url{https://doi.org/10.1007/978-3-642-01539-7_16} 
\bibitem{Blaizot:1986} J. P. Blaizot and L. D. McLerran. Jets in Expanding Quark - Gluon Plasmas. Phys. Rev. D \textbf{34}, 2739 (1986).
\bibitem{Gyulassy:1990} M. Gyulassy and M. Plumer. Jet Quenching in Dense Matter. Phys. Lett. B \textbf{243}, 432 (1990).
\bibitem{Wang:1992} X. N. Wang and M. Gyulassy. Gluon shadowing and jet quenching in A + A collisions at $\sqrt{s_{\rm NN}}$ = 200 GeV. Phys. Rev. Lett. \textbf{68}, 1480 (1992)
\bibitem{Baier:1995} R. Baier, Y. L. Dokshitzer, S. Peigne and D. Schiff. Induced gluon radiation in a QCD medium. Phys. Lett. B \textbf{345}, 277 (1995).
\bibitem{Baier:1997} R. Baier, Y. L. Dokshitzer, A. H. Mueller, S. Peigne, and D. Schiff. Radiative energy loss and $p_T$ broadening of high-energy partons in nuclei.  Nucl. Phys. B \textbf{484}, 265 (1997).
\bibitem{Qin:2015} G. Y. Qin and X. N. Wang. Jet quenching in high-energy heavy-ion collisions. Int. J. Mod. Phys. E \textbf{24}, 1530014 (2015).
\bibitem{Gyulassy:2004} M. Gyulassy, I. Vitev, X.-N. Wang, and B.-W. Zhang. Jet quenching and radiative energy loss in dense nuclear matter. Quark-Gluon Plasma, R. C. Hwa and X.-N. Wang, Eds., p. 123, World Scientific, Singapore, 2004. 


\bibitem{Adcox:2001jp} K. Adcox {\em et al.} (PHENIX Collaboration). Suppression of hadrons with large transverse momentum in central Au+Au collisions at $\sqrt{s_{NN}}$ = 130 GeV.  Phys.Rev. Lett. \textbf{88},  022301 (2002).
\bibitem{Adler:2002tq} C. Adler {\em et al.} (STAR Collaboration). Disappearance of back-to-back high $p_{T}$ hadron correlations in central Au+Au collisions at $\sqrt{s_{NN}}$ = 200 GeV. Phys. Rev. Lett. \textbf{90}, 082302 (2003). 
\bibitem{Adcox:2002pe} K. Adcox {\em et al.} (PHENIX Collaboration). Centrality dependence of the high $p_T$ charged hadron suppression in Au+Au collisions at $\sqrt{s_{\rm NN}}$ = 130 GeV. Phys. Lett. \textbf{ B561}, 82 (2003).
\bibitem{Adler:2003qi} S. S. Adler {\em et al.} (PHENIX Collaboration). Suppressed $\pi^0$ production at large transverse momentum in central Au+ Au collisions at $\sqrt{s_{NN}}$ = 200 GeV. Phys. Rev. Lett. \textbf{91}, 072301 (2003). 
\bibitem{Adams:2003kv} J. Adams {\em et al.} (STAR Collaboration). Transverse momentum and collision energy dependence of high-$p_T$ hadron suppression in Au+Au collisions at ultrarelativistic energies. Phys. Rev. Lett. \textbf{91}, 172302 (2003). 
\bibitem{Adams:2003im} J. Adams {\em et al.} (STAR Collaboration). Evidence from d+Au measurements for final state suppression of high-$p_T$ hadrons in Au+Au collisions at RHIC. Phys. Rev. Lett. \textbf{91}, 072304 (2003).
\bibitem{Back:2003qr} B. Back {\em et al.} (PHOBOS Collaboration). Charged hadron transverse momentum distributions in Au + Au collisions at $\sqrt{s_{\rm NN}}$ = 200 GeV. Phys. Lett. B \textbf{578}, 297 (2004).
\bibitem{Arsene:2003yk} I. Arsene {\em et al.} (BRAHMS Collaboration). Transverse momentum spectra in Au+Au and d+Au collisions at $\sqrt{s_{NN}}=$  200 GeV and the pseudorapidity dependence of high-$p_T$  suppression. Phys. Rev. Lett. \textbf{91}, 072305 (2003). 
\bibitem{Adare:2006nr} A. Adare {\em et al.} (PHENIX Collaboration). System Size and Energy Dependence of Jet-Induced Hadron Pair Correlation Shapes in Cu+Cu and Au+Au Collisions at $\sqrt{s_{NN}}=$ 200 and 62.4 GeV. Phys. Rev. Lett. \textbf{98}, 232302 (2007).	
\bibitem{PHENIX:2005} S. S. Adler {\em et al.} (PHENIX Collaboration). Formation of dense partonic matter in relativistic nucleus-nucleus collisions at RHIC: Experimental evaluation by the PHENIX collaboration. Nucl. Phys. A \textbf{757}, 184 (2005).
\bibitem{STAR:2005} J. Adams {\em et al.}  (STAR Collaboration). Experimental and theoretical challenges in the search for the quark gluon plasma: The STAR Collaboration's critical assessment of the evidence from RHIC collisions. Nucl. Phys. A \textbf{757}, 102 (2005).
\bibitem{PHOBOS:2005} B. Back {\em et al.}  (PHOBOS Collaboration). The PHOBOS perspective on discoveries at RHIC. Nucl. Phys. A \textbf{757}, 28 (2005).
\bibitem{BRAHMS:2005} I. Arsene {\em et al.} (BRAHMS Collaboration). Quark gluon plasma and color glass condensate at RHIC? The Perspective from the BRAHMS experiment. Nucl. Phys. A \textbf{757}, 1 (2005).
\bibitem{Adare:2008cg}
A.~Adare \textit{et al.} (PHENIX Collaboration). Quantitative Constraints on the Opacity of Hot Partonic Matter from Semi-Inclusive Single High Transverse Momentum Pion Suppression in Au+Au collisions at $\sqrt{s_{NN}}=$ 200 GeV. Phys. Rev. C \textbf{77}, 064907 (2008).

\bibitem{Aamodt:2010jd} K. Aamodt {\em et al.}  (ALICE Collaboration). Suppression of Charged Particle Production at Large Transverse Momentum in Central Pb-Pb Collisions at $\sqrt{s_{NN}}$ = 2.76 TeV. Phys. Lett. \textbf{B 696}, 30 (2011).
\bibitem{Aad:2010bu} G. Aad {\em et al.} (ATLAS Collaboration). Observation of a Centrality-Dependent Dijet Asymmetry in Lead-Lead Collisions at $\sqrt{s_{NN}}$ = 2.77 TeV with the ATLAS Detector at the LHC. Phys. Rev. Lett. \textbf{105}, 252303 (2010).
\bibitem{Chatrchyan:2011sx} S. Chatrchyan {\em et al.} (CMS Collaboration). Observation and studies of jet quenching in PbPb collisions at nucleon-nucleon center-of-mass energy = 2.76 TeV. Phys. Rev. C \textbf{84}, 024906 (2011).
\bibitem{Chatrchyan:2011pb} S. Chatrchyan {\em et al.} (CMS Collaboration). Dependence on pseudorapidity and centrality of charged hadron production in PbPb collisions at a nucleon-nucleon centre-of-mass energy of 2.76 TeV. JHEP \textbf{1108}, 141 (2011).
\bibitem{Aamodt:2011vg} K. Aamodt {\em et al.} (ALICE Collaboration). Particle-Yield Modification in Jetlike Azimuthal Dihadron Correlations in Pb-Pb Collisions at $\sqrt{s_{NN}}$ = 2.76 TeV. Phys. Rev. Lett. \textbf{108}, 092301 (2012).
\bibitem{CMS:2012aa} S. Chatrchyan {\em et al.} (CMS Collaboration). Study of high-pT charged particle suppression in PbPb compared to pppp collisions at $\sqrt{s_{NN}}$ = 2.76 TeV.  Eur. Phys. J. \textbf{C 72}, 1945 (2012).
\bibitem{Chatrchyan:2012nia} S. Chatrchyan {\em et al.} (CMS Collaboration). Jet momentum dependence of jet quenching in PbPb collisions at $\sqrt{s_{NN}}$ = 2.76 TeV. Phys. Lett. \textbf{B 712}, 176 (2012).
\bibitem{Chatrchyan:2012gw} S. Chatrchyan {\em et al.} (CMS Collaboration). Measurement of jet fragmentation into charged particles in pp and PbPb collisions at $\sqrt{s_{NN}}$ = 2.76  TeV.  JHEP \textbf{1210}, 087 (2012).
\bibitem{Chatrchyan:2012gt} S. Chatrchyan {\em et al.} (CMS Collaboration). Studies of jet quenching using isolated-photon+jet correlations in PbPb and pp collisions at $\sqrt{s_{NN}}$ = 2.76 TeV. Phys. Lett. \textbf{B 718}, 773 (2013).
\bibitem{Aad:2012vca} G. Aad {\em et al.} (ATLAS Collaboration). Measurement of the jet radius and transverse momentum dependence of inclusive jet suppression in lead-lead collisions at $\sqrt{s_{NN}}$ = 2.76 TeV with the ATLAS detector.  Phys. Lett. \textbf{B 719}, 220 (2013).
\bibitem{Chatrchyan:2013exa} S. Chatrchyan {\em et al.} (CMS Collaboration). Evidence of b-Jet Quenching in PbPb Collisions at $\sqrt{s_{NN}}$ = 2.76 TeV. Phys. Rev. Lett. \textbf{113}, 132301 (2014).
\bibitem{Chatrchyan:2013kwa} S. Chatrchyan {\em et al.} (CMS Collaboration). Modification of Jet Shapes in PbPb Collisions at $\sqrt {s_{NN}}$ = 2.76 TeV. Phys. Lett. \textbf{B 730}, 243 (2014).
\bibitem{Chatrchyan:2014ava} S. Chatrchyan {\em et al.} (CMS Collaboration). Measurement of Jet Fragmentation in PbPb and pp Collisions at $\sqrt{s_{NN}}$ = 2.76 TeV. Phys. Rev. \textbf{C 90}, 024908 (2014).
\bibitem{Aad:2014wha} G. Aad {\em et al.} (ATLAS Collaboration). Measurement of inclusive jet charged-particle fragmentation functions in Pb+Pb collisions at $\sqrt{s_{NN}}$ = 2.76 TeV with the ATLAS detector. Phys. Lett. \textbf{B 739}, 320 (2014).
\bibitem{Aad:2014bxa}  G. Aad {\em et al.}  (ATLAS Collaboration). Measurements of the Nuclear Modification Factor for Jets in Pb+Pb Collisions at $\sqrt{s_{\mathrm{NN}}}$ = 2.76 TeV with the ATLAS Detector. Phys. Rev. Lett. \textbf{114}, 072302 (2015). 

\bibitem{Zakharov:1996} B. Zakharov. Fully quantum treatment of the Landau-Pomeranchuk-Migdal effect in QED and QCD. JETP Lett. {\bf 63}, 952 (1996).
\bibitem{Baier:1997qr} R. Baier, Y. L. Dokshitzer, A.H. Mueller, S. Peigne and D. Schiff. Radiative energy loss of high-energy quarks and gluons in a finite volume quark - gluon plasma. Nucl. Phys. B {\bf 483}, 291 (1997).
\bibitem{Gyulassy:2000} M. Gyulassy, P. Levai and I. Vitev. Jet quenching in thin quark gluon plasmas. 1. Formalism. Nucl. Phys. \textbf{B 571}, 197 (2000).
\bibitem{Gyulassy:2000qr} M. Gyulassy, P. Levai and I. Vitev. NonAbelian energy loss at finite opacity. Phys. Rev. Lett. \textbf{ 85}, 5535 (2000).
\bibitem{Gyulassy:2001er} M. Gyulassy, P. Levai and I. Vitev. Reaction operator approach to nonAbelian energy loss. Nucl.  Phys.  \textbf{B 594}, 371 (2001).
\bibitem{Buzzatti:2011vt} A. Buzzatti and M. Gyulassy. Jet Flavor Tomography of Quark Gluon Plasmas at RHIC and LHC. Phys. Rev.  Lett.  \textbf{ 108}, 022301 (2012).
\bibitem{Guo:2000nz} X. F. Guo and X. N. Wang. Multiple scattering, parton energy loss and modified fragmentation functions in deeply inelastic e A scattering. Phys.  Rev.  Lett.   {\bf 85}, 3591 (2000).
\bibitem{Wang:2001} X.-N. Wang and X.-F. Guo. Multiple parton scattering in nuclei: Parton energy loss. Nucl. Phys. A {\bf 696}, 788 (2001).
\bibitem{Chen:2011vt} X. F. Chen, T. Hirano, E. Wang, X. N. Wang and H. Zhang. Suppression of high $p_{T}$ hadrons in Pb+Pb Collisions at LHC. Phys.  Rev.  C {\bf 84}, 034902 (2011).
\bibitem{Majumder:2009ge} A. Majumder. Hard collinear gluon radiation and multiple scattering in a medium. Phys.  Rev.  D {\bf 85}, 014023 (2012).
\bibitem{Vitev:2009rd} I. Vitev and B. W. Zhang. Jet tomography of high-energy nucleus-nucleus collisions at next-to-leading order. Phys.  Rev.  Lett.   {\bf 104}, 132001 (2010).
\bibitem{Wiedemann:2000} U. A. Wiedemann. Gluon radiation off hard quarks in a nuclear environment: Opacity expansion. Nucl. Phys. \textbf{B 588}, 303 (2000).
\bibitem{Wiedemann:2001}  U. A. Wiedemann. Jet quenching versus jet enhancement: A Quantitative study of the BDMPS-Z gluon radiation spectrum. Nucl. Phys. \textbf{A 690}, 731 (2001).
\bibitem{Arnold:2001} P.B. Arnold, G.D. Moore and L.G. Yaffe. Photon emission from ultrarelativistic plasmas. JHEP {\bf 11}, 057 (2001).
\bibitem{Arnold:2002} P.B. Arnold, G.D. Moore and L.G. Yaffe. Photon and gluon emission in relativistic plasmas. JHEP \textbf{06}, 030 (2002).
\bibitem{Schenke:2009gb} B. Schenke, C. Gale and S. Jeon. MARTINI: An Event generator for relativistic heavy-ion collisions. Phys.  Rev.  C {\bf 80}, 054913 (2009).
\bibitem{Fochler:2010wn} O. Fochler, Z. Xu and C. Greiner. Energy loss in a partonic transport model including bremsstrahlung processes. Phys.  Rev.  C {\bf 82}, 024907 (2010).
\bibitem{He:2015pra} Y. He, T. Luo, X. N. Wang and Y. Zhu. Linear Boltzmann Transport for Jet Propagation in the Quark-Gluon Plasma: Elastic Processes and Medium Recoil. Phys.  Rev.  C {\bf 91}, 054908 (2015), Erratum: [Phys.  Rev. C \textbf{97}, 019902 (2018)].
\bibitem{CasalderreySolana:2007sw} J. Casalderrey-Solana and X. N. Wang. Energy dependence of jet transport parameter and parton saturation in quark-gluon plasma. Phys.  Rev. C \textbf{77}, 024902 (2008).
\bibitem{Burke:2013yra} K. M. Burke {\it et al.} (JET Collaboration). Extracting the jet transport coefficient from jet quenching in high-energy heavy-ion collisions. Phys.  Rev. C \textbf{90}, 014909 (2014).
\bibitem{nestor} N. Armesto {\it et al.}. Percolation approach to quark - gluon plasma and J/$\psi$ suppression. Phys.~Rev.~Lett. {\bf 77}, 3736 (1996).
\bibitem{pajares1} M.~A. Braun and C.~Pajares. Implications of percolation of color strings on multiplicities, correlations and the transverse momentum. Eur. Phys. J. C {\bf 16}, 349 (2000). 
\bibitem{Braun2000} M.~A.~Braun and C.~Pajares. Transverse momentum distributions and their forward backward correlations in the percolating color string approach. Phys.~Rev.~Lett. {\bf 85}, 4864 (2000).
\bibitem{pajares2}M.~A. Braun, F.~del Moral and C.~Pajares. Percolation of strings and the first RHIC data on multiplicity and transverse momentum distributions. Phys.~Rev.~C {\bf 65}, 024907 (2002).
\bibitem{pajares3} J.~{Dias de Deus}, C.~Pajares. Percolation of color sources and critical temperature. Phys.~Lett.~B {\bf 642}, 455 (2006).
\bibitem{Phyreport} M.~A. Braun {\it et al.}. De-Confinement and Clustering of Color Sources in Nuclear Collisions. Phys.~Rep. {\bf 599}, 1 (2015). 
\bibitem{perx} J. Dias de Deus and C. Pajares. String Percolation and the Glasma. Phys.~Lett.~B {\bf 695}, 211 (2011).
\bibitem{Braun2003} M.~A.~Braun, F.~del Moral, and C.~Pajares. Centrality dependence of the multiplicity and transverse momentum distributions at RHIC and LHC and the percolation of strings. Nucl.~Phys.~A {\bf715}, 791 (2003).

\bibitem{schw} J. Schwinger. Gauge Invariance and Mass. 2.. Phys. Rev. {\bf 128}, 2425 (1962).
\bibitem{epjc71} R.~P.~Scharenberg, B.~K.~Srivastava and A.~S.~Hirsch. Percolation of Color Sources and the determination of the Equation of State of the Quark-Gluon Plasma (QGP) produced in central Au-Au collisions at $\sqrt{s_{NN}}$ = 200 GeV. Eur.~Phys.~J.~C {\bf 71}, 1510 (2011).	
\bibitem{Isichenko} M.~B.~Isichenko. Percolation, statistical topography, and transport in random media. Rev.~Mod.~Phys. {\bf 64}, 961 (1992).
\bibitem{Satz2000}H. Satz. Color deconfinement in nuclear collisions. Rep. Prog. Phys. {\bf 63}, 1511 (2000).
\bibitem{aditya:pos2019} A.~N.~Mishra {\it et al.}. ALICE data in the framework of the Color String Percolation Model. {\it PoS } {\bf LHCP2019} (2019) 004.
\bibitem{cunq} L. Cunqueiro, J.~{Dias de Deus}, C.~Pajares. Nuclear like effects in proton-proton collisions at high energy. Eur.~Phys.~J.~C {\bf 65}, 423 (2010). 
\bibitem{andres}C.~Andres, A.~Moscoso and C.~Pajares. Onset of the ridge structure in AA, pA, and pp collisions. Phys.~Rev.~C {\bf 90}, 054902 (2014).  	
\bibitem{cpod13}R.~P. Scharenberg. The QGP Equation of State by Measuring the Color Suppression Factor at RHIC and LHC Energies.  {\it PoS} {\bf CPOD 2013}, 017 (2013).	
\bibitem{eos2}J.~Dias de~Deus {\it et al.}. Clustering of color sources and the shear viscosity of the QGP in heavy ion collisions at RHIC and LHC energies. Eur.~Phys.~J.~C {\bf 72}, 2123 (2012). 	
\bibitem{IS2013} B.~K. Srivastava. Percolation Approach to Initial Stage Effects in High Energy Collisions. Nucl.~Phys.~A {\bf 926}, 142 (2014).	
\bibitem{eos3} J.~Dias de~Deus {\it et al.}. Transport Coefficient to Trace Anomaly in the Clustering of Color Sources Approach. Phys.~Rev.~C {\bf 93}, 024915 (2016).
\bibitem{Sahoo:2018dcz} P.~Sahoo, S.~De, S.~K.~Tiwari and R.~Sahoo. Energy and Centrality Dependent Study of Deconfinement Phase Transition in a Color String Percolation Approach at RHIC Energies. Eur.\ Phys.\ J.\ A {\bf 54}, 136 (2018).
\bibitem{Sahoo:2017umy} P.~Sahoo, S.~K.~Tiwari, S.~De, R.~Sahoo, R.~P.~Scharenberg and B.~K.~Srivastava. Thermodynamic and transport properties in Au + Au collisions at RHIC energies from the clustering of color strings. Mod.\ Phys.\ Lett.\ A {\bf 34}, 1950034 (2019).
\bibitem{Sahoo:2019xjq} P.~Sahoo, R.~Sahoo and S.~K.~Tiwari. Wiedemann-Franz law for hot QCD matter in a color string percolation scenario. Phys. Rev. D \textbf{100}, 051503 (2019).
\bibitem{Sahoo:2018dxn} P.~Sahoo, S.~K.~Tiwari and R.~Sahoo. Electrical conductivity of hot and dense QCD matter created in heavy-ion collisions: A color string percolation approach. Phys. Rev. D \textbf{98}, 054005 (2018).
\bibitem{Sahu:2020nbu} D.~Sahu, S.~Tripathy, R.~Sahoo and S.~K.~Tiwari. Formation of a Perfect Fluid in pp p-Pb, Xe-Xe and Pb-Pb Collisions at the Large Hadron Collider Energies. arXiv:2001.01252 [hep-ph].
\bibitem{Sahu:2020mzo} D.~Sahu and R.~Sahoo. Thermodynamic and transport properties of matter formed in pp, p-Pb, Xe-Xe and Pb-Pb collisions at the Large Hadron Collider using color string percolation model. J. Phys. G \textbf{48}, 125104 (2021).
\bibitem{Mishra:2020epq} A.~N.~Mishra, G.~Pai\'c, C.~Pajares, R.~P.~Scharenberg and B.~K.~Srivastava. Deconfinement and degrees of freedom in pp and A-A collisions at LHC energies.
Eur. Phys. J. A \textbf{57}, 245 (2021).
\bibitem{cgc} L. McLerran and R. Venugopalan. Computing quark and gluon distribution functions for very large nuclei. Phys.~Rev.~D {\bf 49}, 2233 (1994).
	  
\bibitem{alice1} S.~Acharya {\it et al.} (ALICE Collaboration). Charged-particle production as a function of multiplicity and transverse spherocity in pp collisions at $\sqrt{s}$ = 5.02 and 13 TeV. Eur.~Phys.~J.~C {\bf 79}, 857 (2019). 
\bibitem{alice2} S.~Acharya {\it et al.} (ALICE Collaboration). Transverse momentum spectra and nuclear modification factors of charged particles in pp, p-Pb and Pb-Pb collisions at the LHC. JHEP {\bf 11}, 013 (2018).
\bibitem{alice3} J.~Adam {\it et al.} (ALICE Collaboration). Transverse momentum spectra and nuclear modification factors of charged particles in Xe-Xe collisions at $\sqrt{s_{\rm NN}}$ = 5.44 TeV. Phys. Lett. B {\bf 788}, 166 (2019).
\bibitem{cross} L. McLerran,  M. Praszalowicz, and B. Schenke. Transverse Momentum of Protons, Pions and Kaons in High Multiplicity pp and pA Collisions: Evidence for the Color Glass Condensate?. Nucl. Phys. A {\bf 916}, 210 (2013).
\bibitem{glauber} C. Loizides. Glauber modeling of high-energy nuclear collisions at the subnucleon level. Phys.~Rev.~C {\bf 94}, 024914 (2016).
\bibitem{bec1} F. Becattini, P. Castorina, A. Milov  and H. Satz. A Comparative analysis of statistical hadron production. Eur.~Phys.~J.~C {\bf 66}, 377 (2010).

\bibitem{ALICE:2015xmh} J.~Adam \textit{et al.} (ALICE Collaboration). Direct photon production in Pb-Pb collisions at $\sqrt{s_{NN}}$ = 2.76 TeV. Phys. Lett. B \textbf{754}, 235 (2016).

\bibitem{Bjorken} J.~D.~Bjorken. Highly Relativistic Nucleus-Nucleus Collisions: The Central Rapidity Region. Phys.~Rev.~D {\bf 27} 140 (1983).	
\bibitem{Wong} C. Y. Wong, Introduction to High Energy Heavy Ion Collisions (World Scientific, Signapore, 1994). \url{https://doi.org/10.1142/1128}


\bibitem{Baier:2008js} R.~Baier and Y.~Mehtar-Tani. Jet quenching and broadening: The Transport coefficient q-hat in an anisotropic plasma. Phys. Rev. C \textbf{78}, 064906 (2008).
\bibitem{Liu:2006} H. Liu, K. Rajagopal and U. A. Wiedemann. Calculating the jet quenching parameter from AdS/CFT. Phys. Rev. Lett. \textbf{97}, 182301 (2006).
\bibitem{Majumder:2007} A. Majumder, B. Muller and X. N. Wang. Small shear viscosity of a quark-gluon plasma implies strong jet quenching. Phys. Rev. Lett. \textbf{99}, 192301 (2007).
\bibitem{Xu:2013} J. Xu. Shear viscosity of nuclear matter. Nucl. Sci. Tech. \textbf{24}, 50514 (2013).
\bibitem{Baier:2003} R. Baier. Jet quenching. Nucl. Phys. \textbf{A 715}, 209 (2003).
\bibitem{Su:2012} N. Su. A brief overview of hard-thermal-loop perturbation theory. Commun. Theor. Phys. \textbf{57}, 409 (2012).
\bibitem{Baier:1996} R. Baier, Y. L. Dokshitzer, A. H. Mueller, S. Peigne, S. Peigne and D. Schiff. The Landau-Pomeranchuk-Migdal effect in QED. Nucl.Phys. \textbf{B 478}, 577 (1996).
 \end{thebibliography}
\end{document}